\documentclass[12pt]{article}
\usepackage{graphicx}
\usepackage{eufrak}
\usepackage[mathscr]{eucal}
\usepackage{latexsym}
\usepackage{epsfig}
\DeclareGraphicsExtensions{eps,ps}

\newcommand{\beq}{\begin{equation}}
\newcommand{\eeq}{\end{equation}}
\def\be{\begin{equation}}
\def\ee{\end{equation}}
\def\ba{\begin{eqnarray}}
\def\ea{\end{eqnarray}}

\def\kk{\mathfrak{K}}

\def\lL{\mathscr{L}}

\begin{document}
\baselineskip=15.5pt
\pagestyle{plain}
\setcounter{page}{1}

\leftline{\tt hep-th/0106127}

\vskip -1.3cm

\rightline{\small{\tt NUB-3215/Th-01}}
\rightline{\small{\tt HUPT-01/A031}}
\rightline{\small{\tt CTP-MIT-3152}}
\rightline{\small{\tt LA-PLATA-TH-01-06}}


\vspace{.5cm}

\begin{center}

{\Large {\bf Brane Worlds, String Cosmology, and AdS/CFT}}

\vskip 1.5cm
Luis Anchordoqui$^{a,}$\footnote{\tt doqui@hepmail.physics.neu.edu}, 
Jos\'e Edelstein$^{b,}$\footnote{\tt edels@lorentz.harvard.edu},
Carlos Nu\~nez$^{b,}$\footnote{\tt nunez@lorentz.harvard.edu}, 
Santiago Perez Bergliaffa$^{c,}$\footnote{\tt sepb@cbpf.br},\\
Martin Schvellinger$^{d,}$\footnote{\tt martin@ctpbeaker.mit.edu},
Marta Trobo$^{e,}$\footnote{\tt trobo@venus.fisica.unlp.edu.ar}, and
Fabio Zyserman$^{e,}$\footnote{\tt zyserman@venus.fisica.unlp.edu.ar}

\medskip
\medskip
${}^a${\it Department of Physics, Northeastern University\\
Boston, MA 02115, USA}

\medskip
${}^b$
{\it Department of Physics, Harvard University\\
Cambridge, MA  02138, USA}

\medskip
${}^c$
{\it Centro Brasileiro de Pesquisas Fisicas,\\ Rua Xavier Sigaud,
150, CEP 22290-180, Rio de Janeiro, Brazil}

\medskip
${}^d$
{\it Center for Theoretical Physics, Massachusetts Institute of Technology\\
Cambridge, MA 02139, USA}

\medskip
${}^e$
{\it Departamento de F\'{\i}sica, Universidad Nacional de La Plata\\
CC67 La Plata (1900), Argentina}

\vspace{.5cm}

\end{center}

\setcounter{footnote}{0}

\begin{center}
{\large {\bf Abstract}}
\end{center}

\noindent Using the thin-shell formalism we discuss the motion of domain 
walls in de
Sitter and anti-de Sitter (AdS) time-dependent bulks. This motion results in a 
dynamics for the brane scale factor. We show that in the case 
of a clean brane the scale factor describes both singular and non-singular 
universes, with phases of contraction and expansion. These phases can be 
understood as quotients of AdS spacetime by a discrete symmetry group.
We discuss this effect in some detail, and suggest how the AdS/CFT 
correspondence
could be applied to obtain a non perturbative description of brane-world 
string cosmology.

\newpage

\tableofcontents

\section{Introduction}

The intriguing idea that fundamental interactions 
could be understood as manifestations of the existence of 
extra dimensions in our 4-dimensional world can be 
traced back at least to the work of 
Kaluza and Klein (KK) \cite{KK}, with revivals of 
activity that one generically refers to as KK 
theories \cite{renacimiento}.  Over the last two years, a fresh
interest in the topic has been rekindled, mainly due to the realization 
that localization of matter \cite{trapping} and localization of 
gravity \cite{RS} may drastically change the commonly assumed properties 
of such models. 

From the phenomenological perspective the so-called  ``brane worlds''
provide an economic explanation of the hierarchy between the
gravitational and electroweak mass scales.
In the canonical example of Arkani-Hamed, Dimopoulos and Dvali 
(ADD) \cite{arkani}, spacetime 
is a direct product of ordinary 4-dimensional manifold (``our universe'') and a (flat) 
spatial $n$-torus of common linear size $r_c$ and volume $v_n = r_c^n$. 
Of course, Standard Model (SM) fields cannot propagate a large distance 
in the extra dimensions without conflict with observations. 
This is avoided by trapping the fields in a thin shell of 
thickness $\delta \sim M_s^{-1}$ \cite{g}. The only 
particles propagating in the $(4+n)$-dimensional bulk are the 
$(4+n)$ gravitons. Thus,
gravity becomes strong in the entire $(4+n)$-dimensional spacetime at 
a scale $M_s \sim$ a few TeV, which is far below the conventional Planck 
scale, $M_{\rm pl} \sim 10^{18}$ GeV. 
Strictly speaking, the low energy effective 4-dimensional Planck 
scale $M_{\rm pl}$ is related to the fundamental scale of gravity 
$M_*$  via Gauss' Law $M_{\rm pl}^2 = M_*^{2+n} v_n$.
For $n$ extra dimensions one finds that $r_c \approx 10^{30/n -19}$ 
m, assuming $M_* \sim 1$ TeV \cite{susy}. 
This relation immediately suggests that $n=1$ is ruled out, because 
$r_c \sim 10^{11}$ m and, the
gravitational interaction would thus be modified at the scale of our solar
system. However, already for $n=2$,
$r_c \sim$ 1 mm - just the scale where our present day experimental
knowledge about gravity ends \cite{hoyle}. Furthermore,
one can imagine more general scenarios termed asymmetric compactifications, 
where, {\it e.g.}, there are $p$ ``small'' dimensions 
with sizes of $\sim$ 1/TeV
and the effective number of large extra dimensions 
being $n_{\rm eff} = n-p$. Here, the expected number of extra dimensions 
should be 6 or 7 as suggested by string theory \cite{lykken}. 
Naturally, there is a strong 
motivation for immediate phenomenological studies to 
assess the experimental viability of such a radical departure from
previous fundamental particle physics.
Leaving aside table-top experiments and astrophysics (which requires that 
$M_s > 110$ TeV for $n=2$, but only around a few TeV for $n>2$ \cite{astro}), 
there are two ways of probing this scenario. Namely, via the KK--graviton 
emission in scattering processes, or else through the exchange of KK towers 
of gravitons among SM particles \cite{pheno}. The search for extra-dimension 
footprints in collider data has already started. However, as yet no 
observational evidence has been found \cite{exp}.

The ADD scenario is based on the fact that the 4-dimensional 
coordinates are independent of the coordinates of the extra $n$ dimensions. 
Giving up this assumption can lead to a number of other interesting 
models with completely different gravitational behaviors. 
Perhaps the most compelling model along these lines can be built by 
considering a $5$-dimensional anti-de Sitter (AdS) space with a 
single $4$-dimensional boundary \cite{RS}. This boundary is taken 
to be a $p$-brane with intrinsic tension $\sigma$. In the Randall-Sundrum 
(RS) world, there is a bound state of the graviton confined to the brane, 
as well as a continuum of KK modes. At low energies, the bound state 
dominates over the KK states to give an inverse square law if the AdS 
radius is sufficiently small. Therefore, Newton's law is 
recovered on the brane, even with an infinitely large fifth 
dimension. The number of papers discussing variants of this scenario is 
already very large \cite{rs}. Some key papers are \cite{top}. For a 
comprehensive review the reader is referred to \cite{vr}. 

The presence of large extra dimensions modifies the Friedmann equation 
on the brane by the addition of non-linear terms \cite{shiro} which yield new 
cosmological scenarios  \cite{cosmology1}.
Fortunately, a world that undergoes a phase of inflation  
could be long 
lived by a universe like our own at low energies \cite{inflation}.
Specifically, brane-world scenarios are good candidates to describe our 
world as long as the normal rate of expansion has been recovered
by the epoch of nucleosynthesis \cite{after_inflation}. 
We remind the reader that the universe has to be re-heated up to a temperature 
${\cal O}$(MeV) so as to synthethize light elements, because 
the entropy produced during a cold inflationary phase red-shifts away. 
Therefore, any significant departure from 
the standard Friedmann-Robertson-Walker (FRW) scenario could only arise at 
very high energies. A broader study of these ideas is currently 
under way (for an incomplete list of references, see \cite{cosmology2}). 
In this regard, we initiate here the analysis of new inflationary 
brane-worlds that arise from surgically modified evolving spacetimes. 
The dynamics of the bulk in our framework is originated in the symmetries
of the dS and AdS spacetimes, without the need of extra fields in the bulk, 
as in the models discussed in reference \cite{d_bulk}. 

The article is divided in two main parts.
In Section 2 we derive the equation of motion of a 
brane using the thin shell formalism \cite{israel}, in 
which the field equations are re-written as junction conditions relating the 
discontinuity in the brane extrinsic curvature to its vacuum energy.  
Then, we discuss the evolution of a single brane falling into dS and AdS 
spaces. As a 
result of its non interaction with the environment producing the 
gravitational field, the brane tension obeys an internal 
conservation law. Therefore, the motion of the brane can be treated as 
a closed system, or alternatively as a continuous collection of such branes. 
The evolution yields a dynamics for the scale factor. We show that in the 
case of a clean brane ({\it i.e.}, without matter 
in the form of stringy excitations) the scale factor describes both 
singular and non-singular universes, with phases of contraction and expansion.
Section 3 contains a general discussion (and some speculations) on the 
applications of our results in the light of the AdS/CFT correspondence.
Quotients of AdS space by discrete symmetry groups describe  
dynamical AdS bulks. Indeed, it has been suggested \cite{marolf} 
that this is the way in which
AdS/CFT correspondence must be formulated in the case of dynamical spacetimes.
However, since we consider a $p$-brane instead of the AdS boundary itself, for our purposes, the
AdS/CFT correspondence should be considered when gravity is coupled to the conformal theory.

\section{Brane worlds from the connected sum of (A)dS}

\subsection{Field equations}

In spite of the fact that dS space does not seem
to be obtainable from stringy backgrounds, there has been 
a growing interest in dS bulks in recent times \cite{ed}. 
In view of these developments we will discuss the motion of spherical 
branes in both dynamical dS and AdS bulks. 
The following discussion will  refer mostly to branes of dimension 
$d=4$, but the relevant equations will be written for arbitrary $d\geq2$. 

In order to build the class of 
geometries of interest, we
consider two copies of $(d+1)$-dimensional dS (AdS) spaces 
${\cal M}_1$ and ${\cal M}_2$ undergoing 
expansion. Then, we remove from each one identical 
$d$-dimensional regions
$\Omega_1$ and $\Omega_2$ \cite{brandon}. One is left with two 
geodesically incomplete
manifolds with boundaries given by the
hypersurfaces $\partial\Omega_1$ and $\partial \Omega_2$.
Finally, we identify the boundaries up to 
homeomorphism $h: \partial \Omega_1 \rightarrow 
\partial \Omega_2$ \cite{massey}.
Therefore, the resulting manifold that is defined by the connected sum  
${\cal M}_1 \#
{\cal M}_2$, is geodesically complete. The classical action 
of such a system can be cast in the following form
\begin{equation}
S = \frac{L_p^{(3-d)}}{16 \pi} \int_{{\cal M}} d^{d+1}x \sqrt{g} (R - 2 \Lambda)
+ \frac{ L_p^{(3-d)}}{8 \pi} \int_{\partial \Omega} d^dx \sqrt{\gamma} \kk
 +\, \sigma \int_{\partial \Omega} d^dx
\sqrt{\gamma}.
\label{action}
\end{equation}
The first term is the usual
Einstein-Hilbert  action with a cosmological constant
$\Lambda$. The second term is the Gibbons-Hawking
boundary term, necessary for a well-defined variational problem
\cite{GH}.
In our convention, the extrinsic
curvature is defined as $\kk_{MN} =  \nabla_{(M}
\hat{n}_{N)}$,
where $\hat{n}^M$ is the outward
pointing normal vector to the boundary
$\partial \Omega$.\footnote{Throughout the article capital Latin
subscripts run from 1 to $(d+1)$, 
lower Greek subscripts from 1 to $d$, and lower Latin subscripts from
1 to $(d-1)$. As usual,
parenthesis denote symmetrization, $\nabla$ is the 
$(d+1)$-dimensional covariant derivative. 
We adopt geometrodynamic units so that $G\equiv 1$, $c\equiv 1$ and 
$\hbar \equiv L_p^2 \equiv M_p^2$, where $L_p$ and $M_p$ are the Planck 
length and Planck mass, respectively.} 
Here $R$ stands for the $(d+1)$-dimensional Ricci scalar in terms of the
metric $g_{MN}$, while $\gamma$ is
the induced metric on the brane, and $\sigma$ is the brane tension. 

For definiteness, the spatial coordinates on $\partial \Omega$ can be 
taken to be the angular variables $\phi_i$, which for a spherically symmetric 
configuration are always well defined up to an overall rotation.
Generically, the line element of each patch can be written as
\begin{equation}
ds^2 = -dt^2 + A^2(t)\,\,
[\, r^2 \, d\Omega_{(d-1)}^2
+ (1-kr^2)^{-1} \,dr^2 \,],
\label{metrica}
\end{equation}
where $k$ takes the values 1 (-1) for dS (AdS),  
$d\Omega_{(d-1)}$ is the corresponding metric on 
the unit $(d-1)$-dimensional sphere, and $t$ is the proper time of a clock 
carried by any raider of the extra dimension.
The homeomorphism $h$ entails a proper matching of the metric across the
boundary layer \cite{israel}. 
The required junction conditions are most
conveniently derived by introducing Gaussian normal coordinates in the
vicinity of the brane,
\begin{equation}
ds^2 = ({\bf n} \cdot {\bf n})^{-1} d\eta^2
+ \gamma_{\mu\nu}(\eta,x^\mu) dx^\mu
dx^\nu,
\end{equation}
where ${\bf n} = \partial_\eta$, $({\bf n} \cdot {\bf n})$ = 1, (-1) if
$\partial \Omega$ is space-like (time-like). The coordinate $\eta$ 
parameterizes the proper
distance perpendicularly measured from $\partial \Omega$.
Associating negatives values of $\eta$ to one side of the shell
 and positive values to the other side, the energy momentum tensor $T^{MN}$ 
can be written as,
\begin{equation}
T^{MN} (x) = \sigma \,\eta^{\mu\nu}\, \delta_{\mu}^{\phantom{\mu}M} \,
\delta_\nu^{\phantom{\nu}N} \,\delta(\eta) 
+ \frac{3\,\Lambda\,L_p^{(3-d)}}{2\,\pi\,d\,(d-1)} \,g^{MN} 
\,[ \Theta(\eta) + \Theta(-\eta)],
\label{pp}
\end{equation}
where $\Theta$ is the step function, and $\eta^{\mu\nu}$ is the Minkowski metric.
The integration of the field  equations
\begin{equation}
G_M^{\phantom{M}N} = \frac{4 \,\pi}{L_p^{(3-d)}}\, T_M^{\phantom{M}N}
\label{einstein_d+1}
\end{equation}
across the boundary $\partial \Omega$ yields,
\begin{equation}
\Delta \kk_\mu^{\phantom{\mu}\nu} - \delta_\mu^{\phantom{\mu}\nu} \Delta \kk 
= \frac{4 \,\pi}{ L_p^{(3-d)}}\,\, \sigma \,\,
\delta_\mu^{\phantom{\mu}\nu}\,\,,
\label{einstein_d}
\end{equation}
where
\begin{equation}
\Delta \kk_\nu^{\phantom{\nu}\mu} \equiv \lim_{\epsilon\rightarrow 0} 
\kk_\nu^{\phantom{\nu}\mu^{(-)}} -
\kk_\nu^{\phantom{\nu}\mu^{(+)}},
\end{equation}
is the jump in the second fundamental form of the brane
in going from $-\epsilon$ to  $+ \epsilon$ side.
For the case at hand $\Delta \kk_\mu^{\phantom{\mu}\nu} = 
2 \kk_\mu^{\phantom{\mu}\nu^{(-)}}$.

In order to analyze the dynamics of the system, the brane is allowed to move
radially. Let the position of the brane be described by $x^\mu(\tau,\phi_i)
\equiv (t(\tau), a(\tau),\phi_i)$, 
with $\tau$ the proper time (as
measured by co--moving observers on the brane) 
that parameterizes the motion, and the velocity of a piece of 
stress energy at the brane satisfying $u_M u^M =-1$.
Note that, with these assumptions
the  brane will have an ``effective scale factor'' ${\cal A}^2(t) = a^2(t)A^2(t)$. 

The unit normal to the brane reads
\begin{equation}
\hat{n}^M = \left[ \frac{a^\prime\,A}{(1-ka^2)^{1/2}},
0, \dots, 0, \frac{1}{A}[1 - k\,a^2 + (a^\prime\,\,A)^2]^{1/2} \right].
\end{equation}
Using the identity $\partial_\eta = n^\mu \partial_\mu$, plus the standard
relation
\begin{equation}
\kk_\tau^{\phantom{\tau}\tau} = \frac{\partial x^M}{\partial x^\tau}\,\frac
{\partial x^\tau}{\partial x^N}
 \kk_N^{\phantom{N}M},
\end{equation}
the non-trivial components of the extrinsic curvature are given by
\begin{equation}
\kk_{\phi_i}^{\phantom{\phi_i}\phi_i} =
\frac{\dot{A} \, a^\prime}{(1 - k a^2)^{1/2}} +\frac{1}{aA}
[1 - k a^2 + (a^\prime A)^2]^{1/2},
\end{equation}
and
\begin{equation}
\kk_\tau^{\phantom{\tau}\tau}  = 
\frac{a^{\prime\prime} \, A}  {[1-ka^2+(a^\prime
A)^2]^{1/2}} + \frac{2 a^\prime \dot{A}}{[1-ka^2]^{1/2}} + \frac{2\,
a^{\prime 2} \,k\, a \, A} {[1-ka^2+(a^\prime A)^2]^{1/2}\,
[1-ka^2]},
\end{equation}
where $a^\prime \equiv da/d\tau$, and $\dot{A} \equiv dA/dt$. 
The proper time is related to the coordinate time by
\begin{equation}
\frac{d\tau}{dt} = \pm \sqrt{1 - \frac{(A\,\dot{a})^2}{1-ka^2}} \,\,\, .
\label{rel}
\end{equation}
Thus, one can always eliminate one in favor of the other
by using $a^\prime = \dot{a}\,dt/d\tau$. With this constraint  
Eqs. (\ref{einstein_d+1}) and (\ref{einstein_d}) become two 
sets of differential equations relating unknown functions of $t$: $A$, $a$.
In the following subsections we find out particular solutions of this system.

\subsection{dS}

Let us consider two patches of dS spacetimes undergoing expansion. 
This implies setting $A(t) = \ell\, \cosh(t/\ell)$ and $k=1$ in 
Eq.(\ref{metrica}), 
where $\ell^2 = d (d-1)/|\Lambda|$ is the dS radius. 
Fig. 1 shows the Penrose diagram of the model considered in this subsection.

\begin{figure}
\label{0}
\begin{center}
\epsfig{file=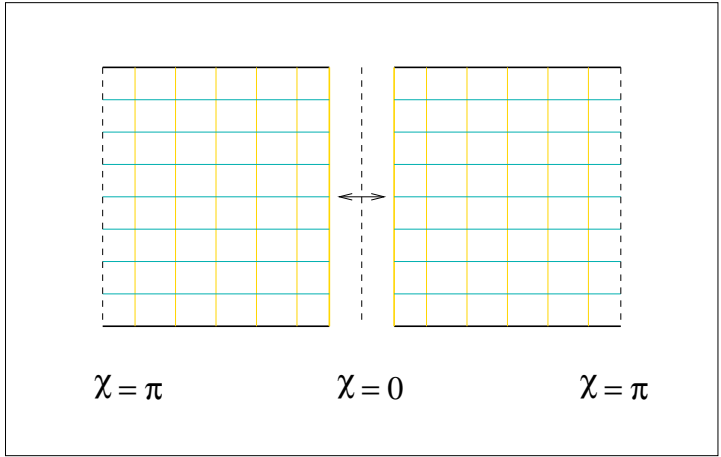,width=10.cm,clip=} 
\caption{{\it Penrose diagram of dS spacetime with a spherical domain wall.
Double arrow stands for identification. Horizontal (vertical) inner
lines are $t$-- ($\chi$--) constant surfaces. 
 The $t=\pm \infty$ surfaces correspond to the top and bottom
horizontal lines, respectively. Vertical dashed lines 
represent the coordinate singularities $\chi=0$ and $\chi=\pi$, typical of 
polar coordinates. We have used the transformation
$r=\sin (\chi)$.}}
\end{center}
\end{figure}

As usual, there is a redundancy between 
the field equations and the covariant conservation of stress-energy. 
A straightforward calculation shows that the equation of motion 
of the brane reads,
\begin{equation}
\frac{4 \,\pi}{L_p^{(3-d)}\,(d-1)}\, \,\sigma = \frac{\pm 
\dot{a}\, \sinh(t/\ell) 
+ [a \,\ell \,\cosh(t/\ell)]^{-1}\,(1 - a^2)}{\{1 - a^2 - 
[\ell\, \cosh(t/\ell)\,
\dot{a}]^2 \}^{1/2}}. 
\label{LAEQ}
\end{equation}
We study first the case $\sigma =0$. 
Integration of Eq.(\ref{LAEQ}) yields,
\begin{equation}
a(t) = \sqrt{{\cal C}\,\tanh^2 (t/\ell) +1},
\label{m1}
\end{equation}
and
\begin{equation}
a(t) = \sqrt{{\cal C}\,\coth^2 (t/\ell) +1},
\label{colon}
\end{equation}
for positive and negative sign in Eq.(\ref{LAEQ}), respectively. ${\cal 
C}$ is an integration constant which 
we will set to 1 in what follows. Note that the proper time is always real for both 
solutions. 

Let us consider first Eq.(\ref{m1}).
A plot of the effective scale factor for this case, Fig.~2(i), 
shows that it corresponds to  
an  eternal non-singular universe 
that undergoes a phase of accelerated contraction up to a minimum volume at 
$t = 0$, 
and then expands again. Note that, for all $t$ the expansion of the bulk 
dominates the cosmological evolution of the brane.
                   
If we define the
Hubble parameters
$H_{\rm brane}= 2\,[\dot{a}/{a} +\dot{A}/{A}]$
and $H_{\rm bulk}=2\,\dot{A}/{A}$, 
we can compare the rates of expansion by plotting
$\Upsilon=H_{\rm brane}/H_{\rm bulk}$. This is shown in Fig.~2(ii). 
For $t\to - \infty$, both the brane 
and the bulk are 
infinitely large and start contracting 
at equal rates. Later, the contraction of the brane is faster than the one
of the bulk, until the minimum is reached. 
The subsequent evolution is the mirror image of the
evolution from $-\infty$ to $0$. The weighted rate of evolution 
(expansion or contraction) is 
larger on the brane for all $t$. 
For $|t| \ll 1$ the rate of expansion on the brane $H_{\rm brane}$ 
is decoupled from that on the 
bulk with a behavior given by $\Upsilon \sim 2 - 2 \,t^2 +10\,t^4/3 + 
{\cal O} (t^5)$. For large values of $t$, $\Upsilon = 
2 \,{\rm cosh}^2 (t) \, {\rm sech} (2t)$. 

Now, we turn to the analysis of the second
solution given by  Eq.(\ref{colon}). As in the previous case, 
both the bulk and the brane have infinite volume
at $t\rightarrow -\infty$, 
see Fig.~3(i). Due to 
the fact that $a(t)$ is finite for 
$\lim_{t \rightarrow \pm \infty}$, 
asymptotically the expansion of the bulk will dominate again 
the evolution of the 
spacetime. However, as 
$t$ approaches to zero, $\coth(t)$ diverges, so that $a(t)$ drives the 
evolution of the brane. 
The behavior of $\Upsilon$ is plotted in Fig.~3(ii). 
This universe is composed of two disconnected branches,
symmetric with respect to $t=0$. The branch on the left starts with an infinite volume,
contracts up to a minimum, and then re-expands back to infinite volume. 
It should be stressed that near $t=0$, a finite interval of the coordinate 
time $t$ is associated to an infinite interval of proper time.

\begin{figure}
\label{1}
\begin{center}
\epsfig{file=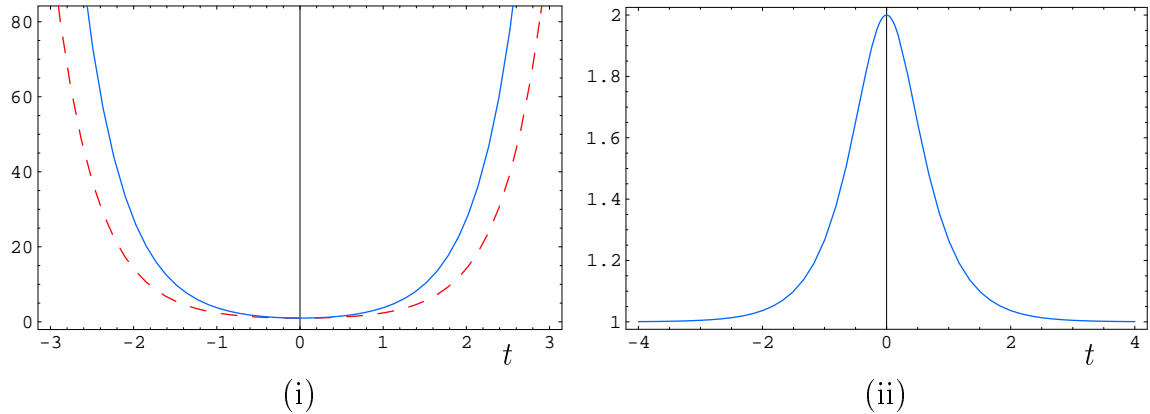,width=16.cm,clip=} 
\caption{{\it (i) Brane (solid-line) and bulk (dashed-line) scale factors.
(ii) Ratio between the brane and bulk expansions, $\Upsilon(t)$. 
We set $\ell =1$ and $\sigma =0$.}}
\end{center}
\end{figure} 

Notice that both solutions exhibit a bounce. They 
can be understood also as a special 
class of wormhole, a Tolman wormhole \cite{tolmanwh}, which entails 
a violation of the strong energy condition on the brane. 

\begin{figure}
\label{2}
\begin{center}
\epsfig{file=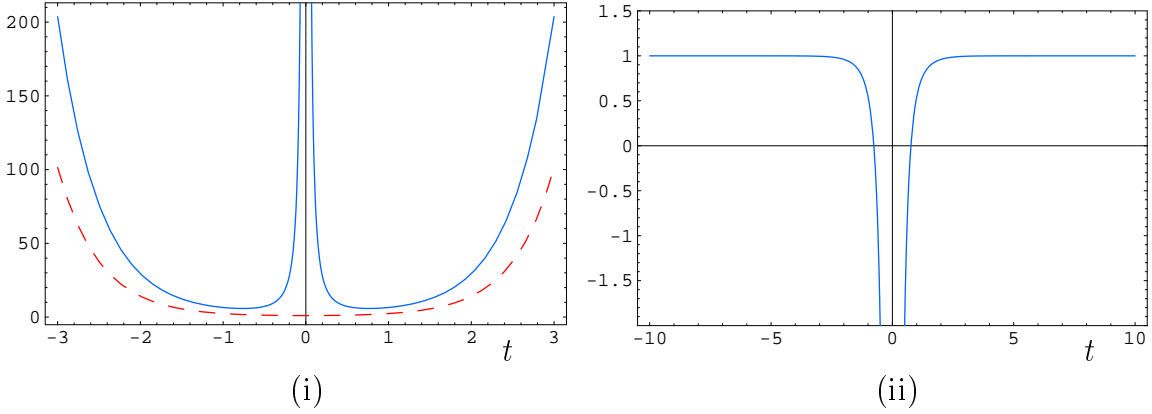,width=16.cm,clip=} 
\caption{{\it (i) Brane (solid-line) and bulk (dashed-line) scale factors.
(ii) Ratio between the brane and bulk expansions, $\Upsilon(t)$. 
We chose $\ell =1$ and $\sigma =0$, as in the previous case.}}
\end{center}
\end{figure} 

In order to obtain a solution for any $\sigma>0$ we must solve 
Eq.(\ref{LAEQ}) numerically.  
For simplicity  we set $\ell =1$. After a brief calculation, Eq.(\ref{LAEQ})
(keeping only the $+$ sign) 
can be casted as
\begin{equation}
A(a,t)\, \dot{a}^2 + B (a,t) \,\dot{a} +C(a,t) = 0,
\label{f-uno}
\end{equation}
where
\begin{eqnarray}
A(a,t) &=& a^2 \, \cosh^2(t)\,  (\overline{\sigma}^2 \, \cosh^2(t) + 
\sinh^2(t)), \\
B(a,t) &=& (1-a^2)\, a \, \sinh(2t),   \\
C(a,t) &=& (1-a^2)^2  - \overline{\sigma}^2 \,\cosh^2(t)\, a^2 \,(1-a^2),
\end{eqnarray}
and the factor $4\, \pi/\,[L_p^{(3-d)}\, (d -1)]$ has been absorbed in
$\overline{\sigma}$. From Eq.(\ref{f-uno}) it is straightforward to write
\begin{equation}
\dot{a} = \frac{-B(a,t) \pm \sqrt{B^2(a,t) - 4A(a,t)C(a,t)}}{2A(a,t)},\quad a(t_0)=a_0.
\label{f-dos}
\end{equation}
It must be noted here that Eq.(\ref{f-dos}) represents two
different initial value problems due to the $\pm$ sign in front
of the square root. From the discussion above, one would wish that 
$a(0)=1$, when $\sigma \rightarrow 0$. However, this does not lead 
to a well defined problem. In order to overcome this step, we take 
$t_0$ close to 0 (positive or negative). This yields $a_0$ near
1 as desired. For the sake of numerical accuracy  \cite{numrec} the right 
hand side of the above equation must be written in a different 
-- but equivalent -- way. Assuming that $a$ is a complex-valued 
function, we get the following
initial value problems
\begin{equation}
\dot{a} = \frac{-B(a,t) - q \sqrt{B^2(a,t)-4A(a,t)C(a,t)}}{2A(a,t)}, \quad 
a(t_0) = a_0,
\label{f-tres}
\end{equation}
and 
\begin{equation}
\dot{a} = \frac{- 2 C(a,t)}{B(a,t) + q \sqrt{B^2(a,t)-4A(a,t)C(a,t)}}, \quad 
a(t_0) = a_0,
\label{f-cuatro}
\end{equation}
where $q=1$ if 
\begin{equation}
{\rm Re}\,\left\{B^*(a,t) \,\sqrt{B^2(a,t)-4A(a,t)C(a,t)}\right\} \ge 0,
\end{equation}
or $q=-1$ otherwise. 

\begin{figure}
\label{f-f1}
\begin{center}
\epsfig{file=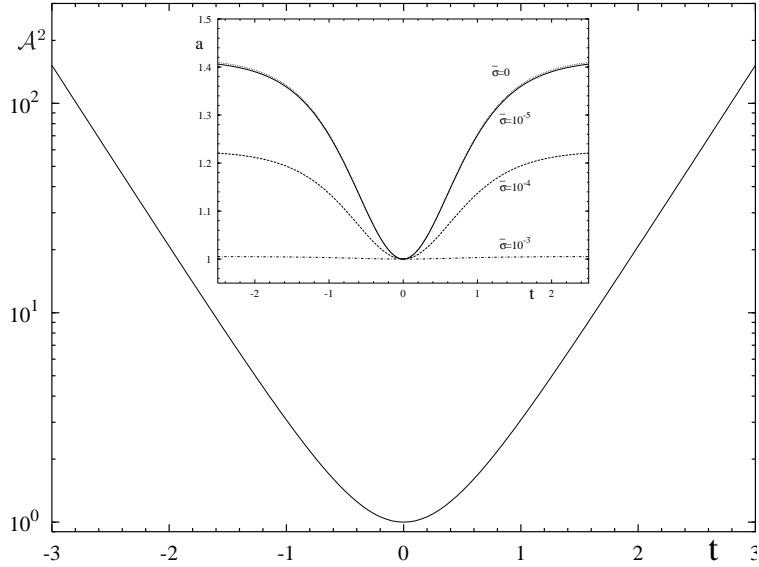,width=10.cm,clip=} 
\caption{{\it The ``effective scale factor'' $\mathcal{A}^2(t)$ is plotted 
for $\overline{\sigma}=10^{-4}$. The embedded figure displays the 
dependence of $a(t)$ with $\overline{\sigma}$ (the imaginary part is zero).}}
\end{center}
\end{figure}

Solutions of Eq.(\ref{f-cuatro}) for different values of 
$\overline{\sigma}$ 
are shown  
in Fig. 4. They were obtained using the well-known 
fourth-order Runge-Kutta method \cite{sb}. In order 
to check the results obtained numerically, we consider the expansion
$a(t)= a_0 + a_1 \, t + {\cal {O}}(t^2)$ around $t=0$. By replacing it
on Eq.(\ref{LAEQ}) we get 
\begin{eqnarray}
a_0^2 &=& \frac{1}{2 \, \overline{\sigma}^2} \,\,
\left( \sqrt{1 + 4 \, \overline{\sigma}^2} -1 \right) \,\,\, ,  
\nonumber \\
a_1^2 &=& 1-2 a_0^2 + a_0^4 \,\,\ .
\end{eqnarray}
For instance, it is straightforward to see that if $\overline{\sigma}=10^{-3}$,
the slope goes to zero.

As in the case $\sigma = 0$, the model describes an eternal 
non-singular universe. By progressively 
increasing the tension of the brane we obtain slower ``cosmological'' 
developments. 
The effective cosmological constant on the brane is given by
\begin{equation}
\lambda_{\rm eff} = \frac{2}{3}\, \frac{d-1}{d} \, \Lambda + 
\frac{1}{8} \,
(d-2)\,(d-1)\,\overline{\sigma}^2.
\end{equation} 
It is interesting to remark that by modifying the vacuum 
energy of the bulk, one can obtain a similar evolution 
for $a$. In other words, by increasing the dS radius $\ell$, $a(t)$ 
becomes smoother as depicted in Fig.~5. Thus, one can say that 
$\Lambda$ and $\sigma$ have opposite effects.

\begin{figure}
\label{4}
\begin{center}
\epsfig{file=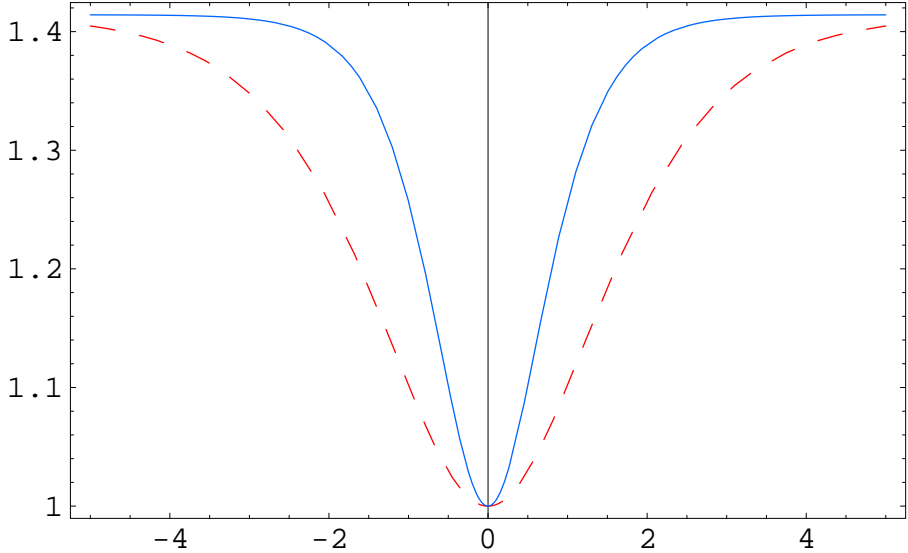,width=12.cm,clip=} 
\caption{{\it Comparison of $a(t)$ for different dS radius. The solid line
stands for $\ell = 1$, whereas the dashed line $\ell =2$. $\sigma$ was set to zero.}}
\end{center}
\end{figure}

At this stage, it is worth noting that the tidal acceleration in the $n^M$ 
direction as measured by observers on the brane with velocity $u^N$ is 
given by $-n_M \, R^M_{\phantom{M}NOP} \,u^N\, n^O\, u^P$,
where $R^M_{\phantom{M}NOP}$ is the Riemann tensor. Recalling that we are 
dealing with a conformally flat bulk, 
\beq
{\rm tidal\; acceleration\; in\; off-brane\; direction} \propto 
\,\Lambda\,\,\,. 
\eeq
Consequently, a positive $\Lambda $  furnishes an acceleration 
away from the brane, giving rise to a non-localized gravity 
model \cite{maartens}. This situation may change when considering models 
with non-local bulk effects (through a nonzero Weyl tensor in the bulk
 \cite{shiro}), or with
the incorporation of moduli 
fields \cite{barcelo-visser} (see Appendix). Note 
that one can compactify the 
spacetime by adding a second brane. In order to do this, one must 
redefine the jump in the 
second fundamental form (in going from $-\epsilon$ to $\epsilon$) between 
the two new disjoint boundaries $\partial \Omega_i \equiv\{r_i =b<a\}$  
as 
\begin{equation}
\Delta \kk_\nu^{\phantom{\nu}\mu} \equiv \lim_{\epsilon\rightarrow 0} 
\kk_\nu^{\phantom{\nu}\mu^{(+)}} -
\kk_\nu^{\phantom{\nu}\mu^{(-)}},
\end{equation}
and then repeat {\it mutatis mutandis} the entire computation \cite{007}.
  
Minkowski or dS compactifications seem to be 
ruled out by some results from low energy string theory \cite{carlos}. 
However, all the ``no go'' arguments rely on the low energy 
limit of type II string theory (or M theory),
where negative tension objects
do not exist. The full string theory could provide objects
such as orientifold planes \cite{polcho}, and compactification to 
RS set ups after including string-loop corrections to the 
gravity action. In addition, massive type II A supergravity could be an 
alternative arena for realizing dS compactifications or RS models on 
manifolds with boundary. Steps in this direction were presented elsewhere 
\cite{tuantran}. A different approach was discussed in \cite{cvetic}.
All in all, if the bulk is conformally flat, then
it is the sign of the bulk cosmological constant that determines 
whether there is gravity trapping or not. Henceforth, we study 
AdS bulks.

\begin{figure}
\label{9}
\begin{center}
\epsfig{file=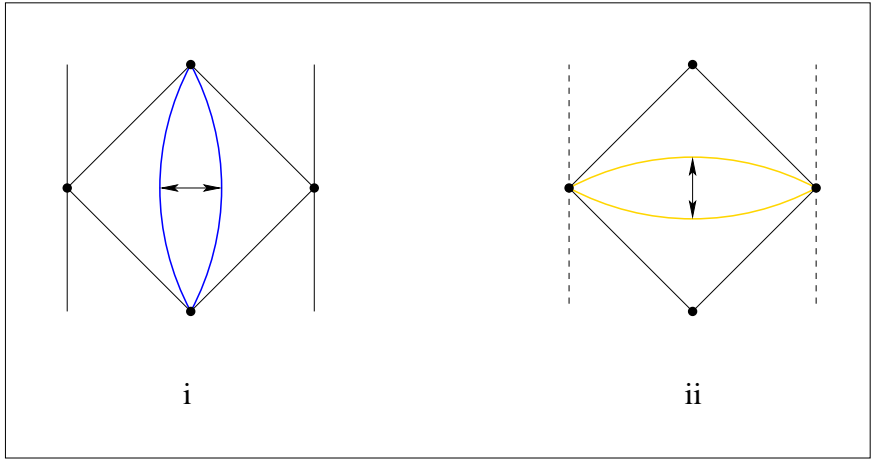,width=12.cm,clip=} 
\caption{{\it (i) Penrose diagram of AdS space with a dS-RS-type 
domain wall. The arrows denote 
identifications. The vertical solid lines represent timelike 
infinity. (ii) AdS 
space with a dS domain wall and pre-surgery regions inside the light-cone.
The vertical dashed-lines 
denote spacelike infinity.}}
\end{center}
\end{figure}

\subsection{AdS}

By setting $A(t) = \sin(t)$, $k=-1$, $\ell =1$ in 
Eq.(\ref{metrica}), it is obtained
\begin{equation}
ds^2 = -dt^2 + \sin^2 (t) \,[ r^2 \,d\Omega_{(d-1)}^2 + 
(1 + r^2)^{-1} \,dr^2] \,\,\, .
\label{ads}
\end{equation}
This is the metric associated to the universal covering space of AdS.
The above metric only covers a part of the universal covering space, even
when all the intervals $t \in [(n\pi, (n+1) \pi]$ (for any
integer $n$) are included. Let us consider only one of the possible patches for
the coordinate $t$ ranged between $0$ to $\pi$. At  $t=\pi$ there is a Cauchy horizon 
connecting to a new spacetime that is physically unreachable.
This horizon is the light-cone emanating from the center of symmetry of the 
solution. Before proceeding further, it is worthwhile to point out that by 
means of the change of variables
$t = i\rho,$ $r = \sinh (i\zeta)$, one obtains the analytic 
continuation of Eq. (\ref{ads}) that reads
\begin{equation}
ds^2=d\rho^2+ \sinh^2 (\rho)\,[d\zeta^2+\sin^2 (\zeta) \,d\Omega_{(d-1)}^2].
\label{euclidmetric}
\end{equation}
In addition, the following coordinate transformations
\begin{equation}
\frac{d\alpha}{d\phi_{(d-1)}}= \tanh (\rho)\,
\sin(\zeta)\,\sin(\phi_i)\,\sin(\phi_j) \dots \, \sin(\phi_{(d-2)}) 
\, \,\, ,\,\,\,\,\,\,
\frac{dy}{d\rho}= (1+ y^2)^{1/2},
\end{equation}
entail a diffeomorphism between the Euclidean AdS-metric given in 
Eq.(\ref{euclidmetric})
and 
\begin{equation}
ds^2 = (1+y^2) d\alpha^2 +
(1+ y^2 )^{-1} dy^2 + y^2 d\Omega_{(d-1)}^2.
\label{euclidinstanton}
\end{equation}
This metric covers the whole spacetime.
Finally, the analytic continuation of 
Eq.(\ref{euclidinstanton}) into real time ($\alpha = iT$) leads to the AdS
region outside the light-cone (Rindler horizon).
It is important to stress that the identification between the two Lorentzian
metrics involves an analytic continuation. Notice that although the 
$(d+1)$-dimensional part 
of the gravity action is boundary independent, the last two terms in 
Eq.(\ref{action}) do indeed depend on the choice of the boundary set up. 
Unlike standard RS-type scenarios \cite{russian-doll}, we proceed here by 
excising two spacetime regions 
internal to the Rindler horizon and gluing the two copies along the 
$(d-1)$-dimensional spheres. The key difference between both scenarios is shown in 
Fig. 6 (more on this below).

\begin{figure}
\label{5}
\begin{center}
\epsfig{file=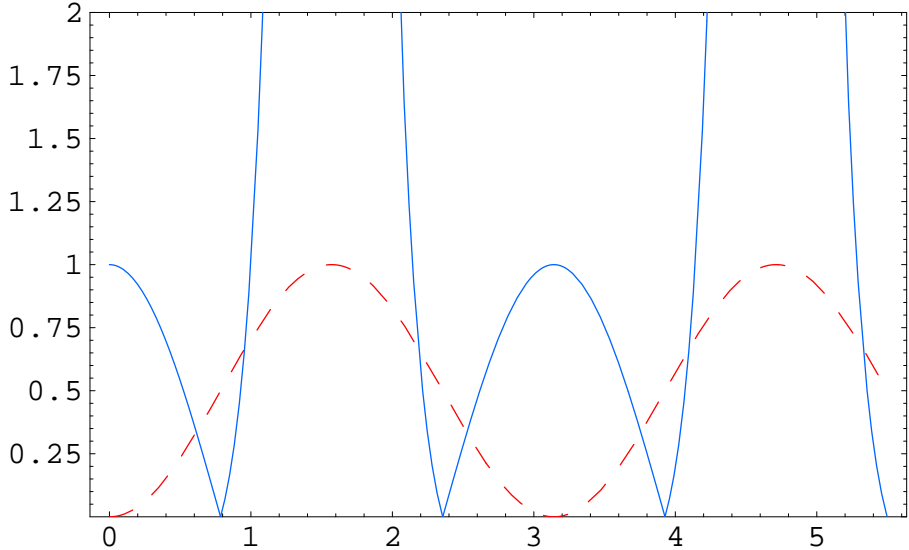,width=12.cm,clip=} 
\caption{{\it Brane (solid-line) and bulk (dashed-line) scale factors as a
function of $t$ with ${\cal C} =1$.}}
\end{center}
\end{figure}

\begin{figure}
\label{6}
\begin{center}
\epsfig{file=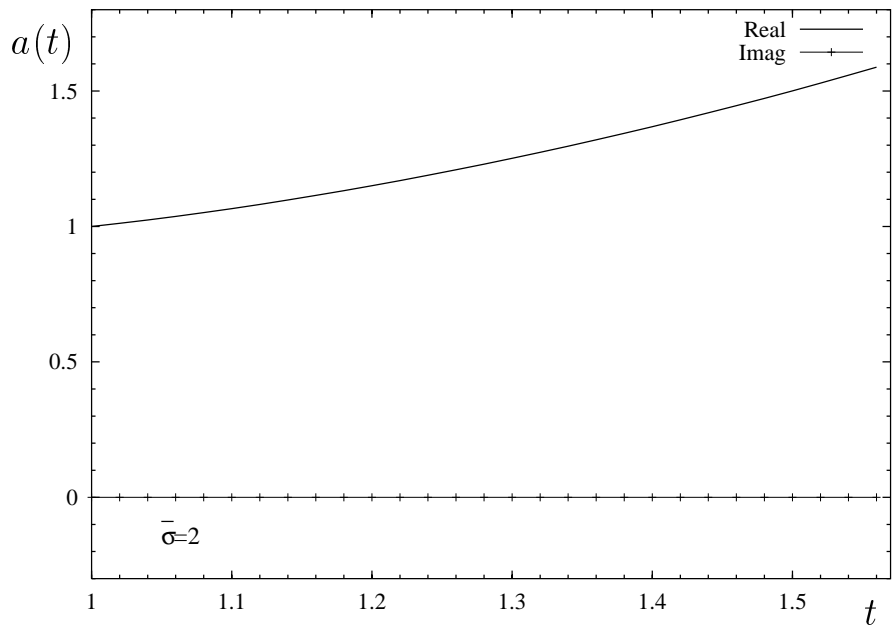,width=12.cm,clip=} 
\caption{{\it Real and imaginary parts of $a(t)$ 
for $\overline{\sigma} =2$.}}
\end{center}
\end{figure}

\begin{figure}
\label{7}
\begin{center}
\epsfig{file=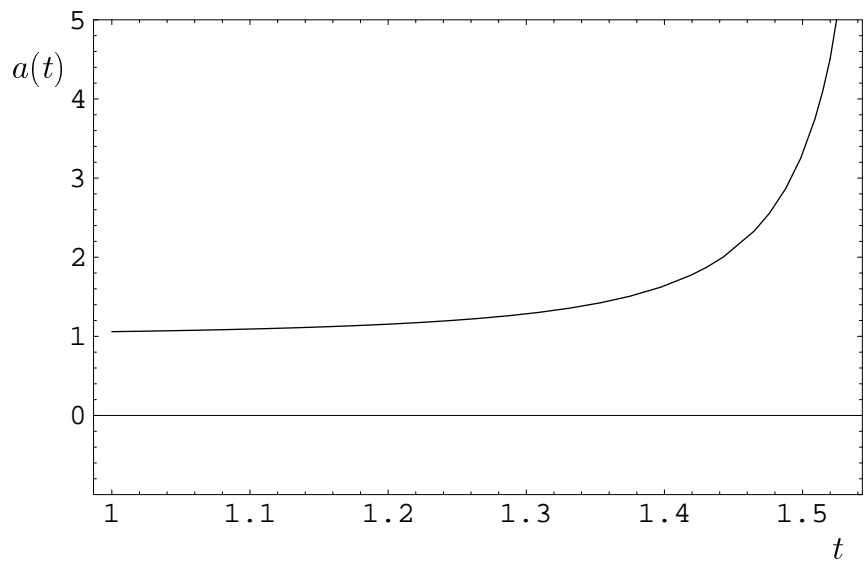,width=12.cm,clip=} 
\caption{{\it Analytical solution for $\overline{\sigma}=0$
from Eq.(\ref{mm}) for $t \in (\pi/4,\pi/2]$. We have considered
${\cal C} = 5 \times 10^{-2}$, so as to approximately reproduce
the initial condition of the numerical solution depicted in Fig. 8.}}
\end{center}
\end{figure} 

With this in mind, the equation of motion of a clean brane sweeping AdS 
is given by  
\begin{equation}
\frac{4 \,\pi  \, \sigma}{L_p^{(3-d)} \, (d-1)} = \frac{\pm\dot{a}\, \cos(t) + [a
\, \sin(t)]^{-1}\,(1 + a^2)}{\{1 + a^2 - [\sin(t)\,\dot{a}]^2
\}^{1/2}}\,\,\,. \label{LAEQ2}
\end{equation} 

For $\sigma = 0$, straightforward integration yields,
\begin{equation}
a(t)= \left\{
\begin{array} {ll} \sqrt{{\cal C}\, \cot^2 (t) -1}  & 
{\rm for} \,\,t \in(0,\pi/4] \\ 
 & \nonumber \\
     \sqrt{{\cal C}\,\tan^2 (t) -1} & {\rm for}\,\, t \in(\pi/4,\pi/2].
\end{array}
\right.
\label{mm}
\end{equation}
To obtain this solution we have exchanged the $\pm$ sign in the 
equation of motion at $t=\pi/4$. In Fig. 7 we show the evolution of 
the bulk and brane scale factors for $t \in [0, 7/4 \pi]$.
Note however, that Eq.(\ref{mm}) does not describe a physical solution 
because the proper time is not real. By checking the Jacobian in
Eq.(\ref{rel}) one can immediately 
realize that the intervals where the proper time
is real are shifted in $\pi/2$ with respect to the corresponding solutions of
Eq.(\ref{mm}). To understand what is going on, let 
us re-write the effective cosmological constant on the brane,
\begin{equation}
\lambda_{\rm eff} = (d-1) \left[\frac{1}{8} (d-2) \overline{\sigma}^2 
- \frac{2}{3} \, \frac{(d-1)}{\ell^2} \right].
\end{equation}
It is easily seen that in order to obtain $\lambda_{\rm eff} > 0$ 
(condition that arises from imposing that the sections $t=$ constant and $a=$ constant 
are $(d-1)$ spheres), we must impose
\begin{equation}
\overline{\sigma}^2 > \frac{16}{3} \, \frac{(d-1)}{(d-2)} \frac{1}{\ell^2}.
\end{equation}  
Thus, the existence of a well behaved solution of a spherical 
domain wall sweeping the internal region to the Rindler horizon 
has an explicit dependence on the dimension of the spacetime through 
the brane tension. 

In order to find the solution for $\overline{\sigma} > 0$ 
we follow the procedure sketched before for the case
of a dS bulk. In Fig. 8 we show the solution of Eq.(\ref{LAEQ2})
(considering again only the plus sign) for $\overline{\sigma}=2$.
It has been obtained by setting $\ell =1$, and with the initial
condition $a(1)=1$. For comparison in Fig. 9 we display the corresponding
analytical solution of Eq.(\ref{mm}), for $t \in (\pi/4,\pi/2]$. 
One can check by inspection that $a(t)$
becomes smoother with increasing $\overline{\sigma}$. This implies that the denominator on the r.h.s. of 
Eq.(\ref{LAEQ2}) is real, rendering a well defined proper time.
Putting all this together, a non-vanishing brane tension
leads to the suitable shifting on $a(t)$.

\section{Non-perturbative string cosmology via AdS/CFT}

A, seemingly different, but in fact closely related subject
we will discuss in this section is the AdS/CFT
correspondence \cite{malda}. This map provides a ``holographic'' projection
of string theory (or M theory) in AdS space, to a conformal field theory 
(CFT) living on its boundary \cite{example}. Actually, more  
general spaces with certain smooth restrictions would typically lead 
to non conformal field theories on their boundary. For asymptotically
AdS spaces the bulk excitations do indeed 
have a correspondent state/operator in the boundary \cite{witten-pol}. 
The duality between the ``strongly coupled gauge theory/weakly coupled 
gravity'' is the face-off of the well known computation of black hole 
quantities via a field theory \cite{strominger-vafa} that naturally 
yields a non-perturbative stringy background. In the standard non-compact 
AdS/CFT set up, gravity is decoupled from the dual boundary theory. However, 
any RS-like model should properly be viewed as a coupling of gravity to 
whatever strongly coupled conformal theory the AdS  geometry is dual 
to \cite{rs_ads/cft}. The holographic description has been recently 
invoked to discuss phenomenological and gravitational aspects of 
RS-models \cite{all}. Here, we try to take advantage of this duality to 
describe cosmological set ups.

There exist a ``new lore'' that convinces us that if our universe is 
five-dimensional, 
it should have evolved to the present situation from M 
theory \cite{Mc}. If this is the case, it is reasonable to describe 
the  evolution of the AdS$_5$ sector (up to stabilization at some scale) by 
\begin{equation}
ds^2 = -dt^2 + \sin^2 (t) \,[ \sinh^2(\zeta) \,d\Omega_{(d-1)}^2+ \,d\zeta^2]. 
\label{ucs_ads}
\end{equation}
In other words, the metric in Eq.(\ref{ucs_ads}) with $SO(d,1)$ isometries 
(which is in fact a subgroup of the full $SO(d,2)$ symmetry group of
AdS) is expected to characterize the dynamics of the system that has 
``stationary'' phases governed by
\begin{equation}
ds^2 = - (1+y^2) \,dT^2 +
(1+ y^2 )^{-1} \, dy^2 + y^2 d\Omega_{(d-1)}^2 \,\,\, .
\label{poi}
\end{equation}
Recall that we set $\ell =1$, and the discussion 
refers to $d=4$. This overall picture is actually related to the notion 
of black holes threaded with collapsing matter, stabilized as static 
objects. Particularly, in such a limit the dual non-perturbative
description for maximally extended Schwarzschild-AdS spacetimes has
been recently put forward \cite{malda_new}.  Applying Maldacena's conjecture 
in the whole dynamical scenario, however, is not 
straightforward. Basically, because the well known correspondence between 
correlation functions in a field theory and string theory backgrounds with 
AdS subspace \cite{witten-pol} does  not have a clear counterpart. We are not going 
to present here a prescription for such a generalization, which is beyond the 
scope of the present article. Instead, we content 
ourselves assuming the existence of a subset of operators/states  
satisfying certain discrete symmetries, and sketch a CFT dual of the dynamical 
gravitational system discussed in the previous section.

In order to do so, we should first draw the reader's attention to some 
generic features of the metrics in Eqs.(\ref{ucs_ads}) and (\ref{poi}). 
The $(p+2)$-dimensional AdS space, can be obtained by taking a hyperboloid 
(for instance see \cite{review})
\begin{equation}
-x_0^2 - x_{p+2}^2 + \sum_{j=1}^{p+1} x_j^2 = -1,
\label{enzo}
\end{equation}
embedded in $(p+3)$-dimensional space with metric
\begin{equation}
ds^2 = -dx_0^2 - dx_{p+2} + \sum_{j=1}^{p+1} dx_j^2.
\end{equation}
By construction, the spacetime 
contains ``anomalies'' in the form of closed time-like curves, 
corresponding to the $S^1$ sector of the hyperboloid. However, by 
unwrapping that circle one can eliminate the causal ``anomalies'', and 
obtain the so called ``universal covering space of AdS''. A convenient
parametrization sets $x_{p+2} = \cos (t)$, so that a constant $t$ surface 
is a constant negative curvature hyperboloid of radius $\sin(t)$, with 
the metric given in Eq.(\ref{ucs_ads}).  One can alternative solve 
Eq.(\ref{enzo})  by setting
\begin{equation}
x_0 =  \cosh(y) \,\sin(T),\,\,\,\,\,\, x_{p+2} =  \cosh(y)\,\cos(T), \,\,\,\,
\,\, x_j = \sinh(y) \,\Omega_j, 
\end{equation}
where $j= 1, \dots, p+1$, and 
$\sum_j \Omega_j^2 =1.$ By inspection of Fig. 10, 
it is easily seen that the universal AdS space 
is conformal  to half of the Einstein static 
universe. While the coordinate system
$(t, \zeta, \phi_i)$, with apparent singularities at $t=n\pi$ ($n \in Z$),
covers only diamond-shaped 
regions, coordinates $(T, y, \phi_i)$ cover the whole space. 
The surface corresponding 
to $t=-\infty$ (solid line) emanates
from $T=0$ bouncing at $T=\pi/2$ towards $T=\pi$. After reflection, the 
solid line represents a hypersurface at $t=\infty$. The vertical 
thin solid lines (with end points $T=0$, $T=\pi$) stand for 
hypersurfaces $y=$ constant, whereas the corresponding horizontal lines 
are $T$= constant curves. Notice that every timelike geodesic emanating 
from any point in the space (to either pass or future) reconverges to an 
image point, diverging again to refocus at a second image point, and so 
on \cite{he}. Therefore, the sets of points which can be reached by future 
directed 
timelike lines starting at a point $p$, is the set of points lying 
beyond the future null cone of $p$, {\it i.e.}, the infinite chain of diamond 
shaped regions similar to the one characterized by $(t, \zeta, \phi_i)$.
Since the total space is non-singular, it should be expected
that the other regions that are outside the diamond-shaped domain can be
included. All points in the Cauchy development of the surface $T=0$
can be reached by a unique geodesic normal to the surface, whereas points 
lying outside the Cauchy development cannot be reached by this kind of 
geodesics.

\begin{figure}
\label{8}
\begin{center}
\epsfig{file=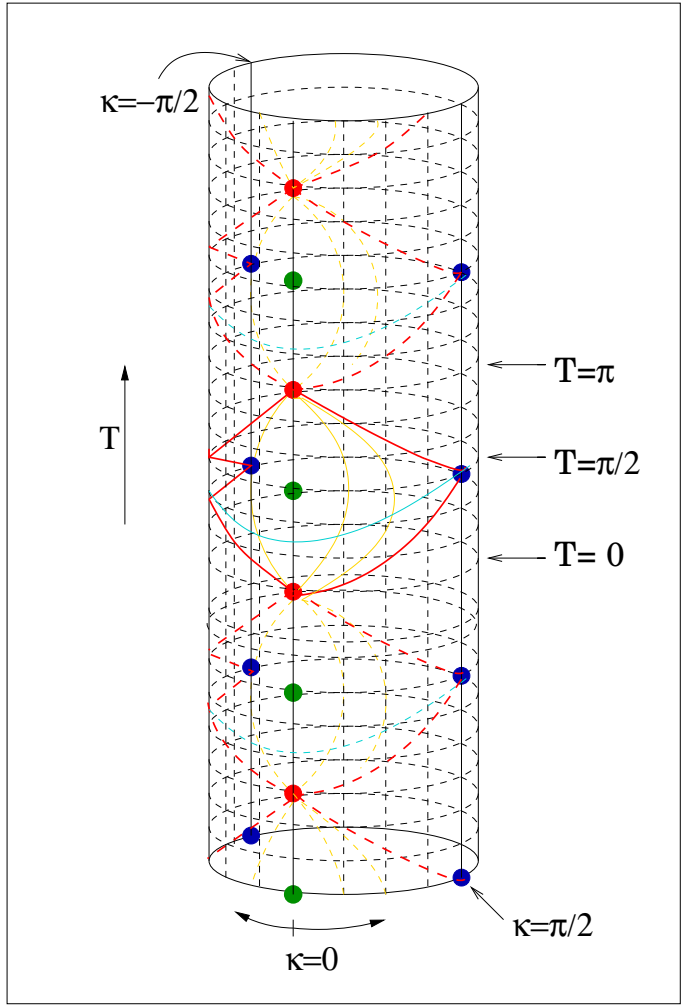,width=9.5cm,clip=} 
\caption{{\it A chain of diamond-shaped regions of the universal covering
of the AdS. Here, $\kappa = 2 \,{\rm arctan} [\exp (\rho)]  - \, \pi/2$.}}
\end{center}
\end{figure}

The infinitely sequence of diamond-shaped regions can also 
be understood as quotients with respect to the universal covering space 
\cite{marolf}. If we consider a quotient of the form  $H_d/\Gamma$ (with a
discrete group of invariances $\Gamma$) such that we get a finite volume
region out of $H_d$, we could interpret this as a cosmological model with
a big bang and a big crunch \cite{btz}.
This cosmological model is a quotient space constructed by identification
of points in  FRW space under the action of a discrete subgroup
$\Gamma$ of the manifold's isometry group $G$. Intuitively, we 
divide the space in different regions such that for any observer placed at
$x_0$, the so-called Dirichlet domain is defined as  
the set of all points closer to $x_0$
than to $\tilde\gamma \,x_0$, being $\tilde\gamma$ an element of the 
group $\Gamma$. Notice that the action of $\tilde\gamma$'s shifts the 
outskirts 
of each point in a given diamond-shaped region to another diamond-shaped 
domain. With this in mind, one can describe the universe we live in by a 
Dirichlet domain. 
Then, we can see the importance of the Cauchy surface mentioned 
above at $T=0$ in
Eq.(\ref{ucs_ads}), since after imposing initial data on this surface
we can cover the complete AdS space.

A particular example of this sort of phenomenon occurs in the extended 
RS-universe discussed in Section 2. The envisaged surgery, shown in Fig. 11, 
starts with two asymptotic AdS regions, each one with two outer and inner 
horizons, that are then repeated an infinite number of times in the maximal 
analytic extension. The brane expansion starts out in the downmost AdS 
region, falling through the future outer horizon to the next incarnation 
of the universe. This process occurs in a finite time $t$. Since we have 
an AdS space and a quotient on it, one can naturally ask,
what is the effect of this quotient on the CFT. This question is of great
interest since {\it the resulting CFT would be a non-perturbative description
of string cosmology}. The idea is to construct a CFT with a Hilbert space
invariant under $\Gamma$. These states would in principle describe
linearized gravity modes in the cosmological background \cite{o}.
The main problem with this construction is that finding such a CFT would
involve to take a quotient on the space on which the CFT lives. The
interesting point discovered in \cite{marolf} is that one can take a
quotient on the operator and state spaces leading to a subset invariant
under $\Gamma$. Before proceeding further, it is instructive to recall 
that instead of a CFT living in a flat Minkowski space, 
we handle a gauge theory coupled to gravity on the brane.

\begin{figure}
\label{10}
\begin{center}
\epsfig{file=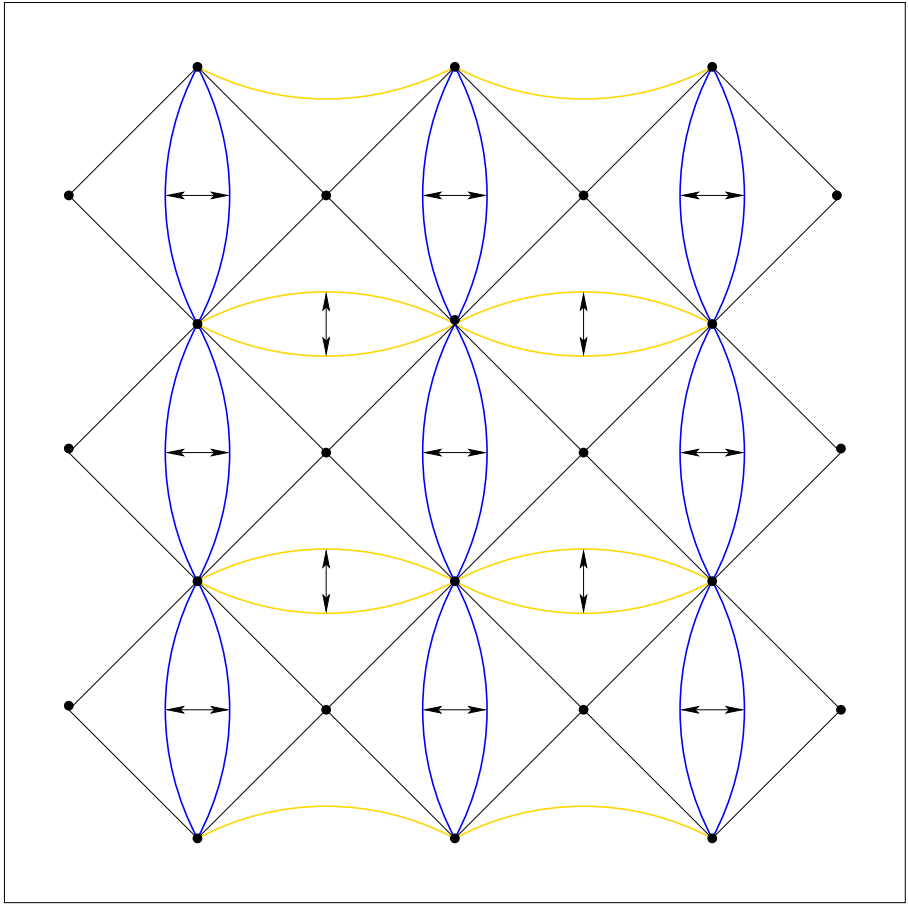,width=12.5cm,clip=} 
\caption{{\it Maximally extended Penrose diagram. 
The past and future RS horizons, are replaced by the past and future 
light-cones obtained after analytical continuation.}}
\end{center}
\end{figure} 

When considering the near horizon geometry 
of D3 branes, AdS$_5 \times S^5$, one deals
with a four dimensional ${\cal {N}}=4$ supersymmetric Yang Mills 
theory as its holographic dual. The symmetries
on the brane side, $SL(2,Z)$ and $SO(6)$
are present on the gravity side as type IIB symmetries. An enormous amount 
of tests on the above relation have been done. Many of them involve 
correlation functions.
Others refer to the matching of the weight of chiral primaries on the   
CFT side with
the masses of the KK modes on the compact part of the geometry.
In the case of M theory, we deal with $AdS_{4} \times S^{7}$ 
(and also with $AdS_{7} \times S^{4}$)
and 3-dimensional  supersymmetric CFT with 16 charges or the (0,2) 
little string theory in each case (they correspond to M2/M5 branes).
In these cases the existence of a matching
between chiral primaries and KK states was also checked in detail. 
The $AdS_6 \times S^4$ (that appears when we consider the D4-D8
system in massive IIA theory \cite{tuantran}) 
leads to five-dimensional CFT. Again, the 
same matching and correspondence was achieved. 
Besides, there exits compactifications of M theory with an AdS$_5$ sector 
(for instance AdS$_5 \times H_2 \times S^4$ \cite{aclaracion}),
where all the previous mentioned features
work in pretty much the same way. The subtlety of these compactifications 
is that they
lead to 4-dimensional CFT's with 4 or 8 supercharges, thus making 
closer contact with the supersymmetric SM-like theories. 

In order to describe a method that can hopefully be applied 
to any of the products discussed above, let us restrict to the case of 
AdS$_3 \times S^3\times T^4$ (the background geometry that appears 
considering D1-D5 system). The gravity system has a dual description 
in terms of a 2-dimensional supersymmetric CFT, with ${\cal {N}}=(4,4)$ supersymmetry. 
The isometries, given by
$SL(2,R)\times SL(2,R)$, coincide with the symmetry generated on the
2-dimensional boundary by
the Virasoro generators $L_{0,\pm},\; \bar{L}_{0,\pm}$. A scalar field
satisfying the Klein-Gordon equation in AdS$_3$, will have eigenmodes in
correspondence with the masses of the chiral primaries on the field
theory side. These states, like all their descendants, are protected
and in correspondence with the KK states on the $S^3 \times T^4$ of
the geometry. Once we have applied the quotient operation, the proposed
way to obtain the correspondence is lifting 
the gravity mode to the uncompactified AdS space. This is carried out by 
considering 
a periodic function defined on each Dirichlet domain, covering the complete
space. The correspondence is between the cosmological gravity mode, and the
state given by the sum of individual states on each domain (on the CFT side). 
It is
important to point out that this sum is not convergent. Therefore, methods
dealing with rigged Hilbert spaces have been applied to define the sum
(the CFT state) properly. 
In addition, as it has been point out in \cite{marolf} the quotients required
to obtain the cosmology break down all the supersymmetries.
One can understand this considering that
a spacetime where some of the supersymmetries remain unbroken 
must be stationary (or sort of
``null-stationary''). However, our quotient spacetime has no Killing fields 
whatsoever.  Note that such a Killing field
would have to be an element of the AdS symmetry group that commutes with the
group we
used to take the quotient.  However, there are not such elements.
Another way of saying this is that, after the quotient,
one will not be able to return (globally) to the static coordinate system
\cite{marolf1}.
It implies that there are not protected quantities in the cosmological scheme.
Other ways of doing the quotient could
in principle, be find. This opens the possibility of different definitions
of this theory in the boundary, which would be very interesting to explore
further. Moreover, it is also important to address
whether
there could exist scenarios in which the above mentioned quotients describing the cosmological evolution do preserve some of the supersymmetries.
We hope to return to these topics in future publications.

\section*{Acknowledgements}

We have benefitted from discussions with Juan Maldacena, 
Stephan Stieberger, and Mike Vaughn. 
We are grateful to Gia Dvali and Tetsuya Shiromizu 
for a critical reading of the 
manuscript and valuable comments. We would like to especially thank Gary 
Horowitz and Don Marolf for sharing with us their expertise on this subject 
and teaching us about their work. This article was partially 
supported by CONICET (Argentina), FAPERj (Brazil), Fundaci\'on Antorchas,
IFLP, the Mathematics Department of the UNLP, the National Science 
Foundation (U.S.), and the U.S. Department of Energy (D.O.E.) under 
cooperative research agreement $\#$DF-FC02-94ER40818.

\newpage

\appendix

\section{Appendix: Flatland dynamics}

In this appendix we will briefly discuss the main features of a novel 
scenario which contains fields in the bulk with Lagrangian $\lL$. 
We start by considering a 
rather general action,
\begin{equation}
S = \frac{L_p^{3-d}}{16 \pi} \int_{{\cal M}} d^{d+1}x \sqrt{g} R  
+ \frac{L_p^{3-d}}{8 \pi} \int_{\partial \Omega} d^dx \sqrt{\gamma} \kk +
\int_{{\cal M}} d^{d+1}x \sqrt{g} \lL + \sigma \int_{\partial 
\Omega} d^dx \sqrt{\gamma}\,\,.
\label{ac2}
\end{equation} 
The standard approach to solve the equations of motion would be to assume some
specific type of fields descibed by $\lL$, and from the physics of 
that source to derive an equation of
state, which together with the field equations 
would determine the behavior of the $(d+1)$-dimensional scale factor. 
In deriving our model, however, the philosophy of solving Einstein's equation 
must be altered somewhat from the usual one: We concentrate on
the geometrical properties of the spacetime, without specifying
$\lL$. 
We are interested here in $(d+1)$-dimensional Ricci-flat spacetime 
undergoing expansion. Thus, we set $k=0$ in Eq.(\ref{metrica}) and we 
replace the  metric in the field equations to obtain the curvature constraint 
\begin{equation}
R = 2 \,d\, \frac{\ddot{A}}{A} + d\,(d-1)\, \frac{\dot{A}^2}{A^2}.
\end{equation} 

\begin{figure}
\label{min}
\begin{center}
\epsfig{file=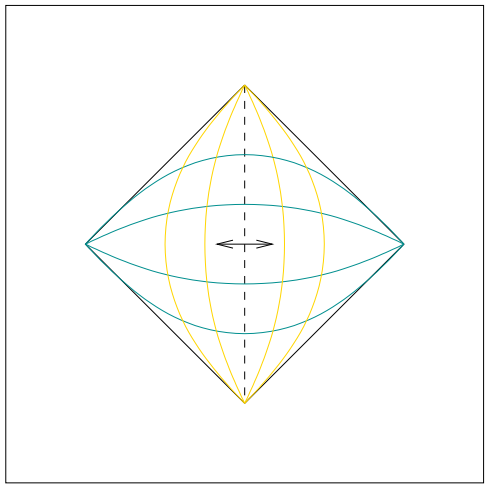,width=7.cm,clip=} 
\caption{{\it Penrose diagram of Minkowski spacetime with a 
spherical domain wall.
Double arrow stands for identification. Horizontal (vertical) inner
lines are $t$-- ($r$--) constant surfaces. 
 The $t=\pm \infty$ surfaces correspond to the top and bottom
horizontal lines, respectively. Vertical dashed line 
represents the coordinate singularities $r=0$, that
occur with polar coordinates.}}
\end{center}
\end{figure} 

\begin{figure}
\label{flatland}
\begin{center}
\epsfig{file=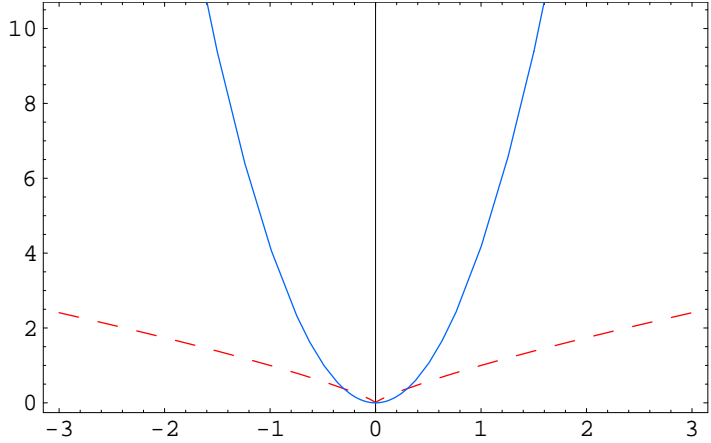,width=9.5cm,clip=} 
\caption{{\it Brane (solid-line) and bulk (dashed-line) scale factors,
for $\sigma=0$. }}
\end{center}
\end{figure}

Now, it is easily seen that  
\begin{equation}
A(t) = \left(\frac{{\cal {K}}_0}{2}\right)^\alpha \,
       [(1+d) (t-{\cal {K}}_1)]^\alpha
\end{equation}
describes a flat solution undergoing expansion, with equation 
of state (in terms of the bulk energy density $\rho$ and pressure $p$),
\begin{equation}
d\, \alpha \,(\alpha-1)\,t^{-2}  = - \frac{4 \pi}{L_p^{3-d}}\, (\rho +d\,p)\,, 
\end{equation}
provided that $\alpha =2/(d+1)$. For simplicity, we have set the integration 
constants ${\cal K}_0 = \alpha$, ${\cal K}_1 = 0$.
Note that the fields 
threading the spacetime violate the $(d+1)$ strong energy condition. 
After recalling that we are dealing with a clean brane and a 
conformally flat bulk, it is straightforward to obtain the equation of motion of a 
brane in this background,
\begin{equation}
\frac{4\pi}{L_p^{3-d}\,(d-1)} \, \sigma = \frac{\pm \dot{a}\, 
\alpha \,t^{\alpha-1} + (a\,\,t^\alpha)^{-1}}{[1- (\dot{a}\,\, t^{\alpha})^2]^{1/2}}\,\,.
\end{equation}
Figure 12 shows the Penrose diagram for this brane-surgery. 
For the case $\sigma =0$, $d=4$ the solution reads,
\begin{equation}
a(t) = \pm \frac{5}{\sqrt{6}} \,t^{3/5} \,\,,
\end{equation} 
where $a_0$ is an integration constant. In Fig. 13, we show the evolution of
$A(t)$ and ${\cal A}(t)$ for $a_0 =0$. The solution (brane and bulk scales 
factor) describes to separate symmetric patches of contraction and expansion 
with a shrinkage down to zero at $t=0$. From Eq.(\ref{rel}) one can easily 
see that the solution does not represent a physical system because 
$(A \dot{a})^2 > 1$. 

It is interesting to point out that in a most general action, the last 
world-volume term in Eq.(\ref{ac2}) would depend on the 
$d$-dimensional Ricci scalar. Such term can not be 
excluded by any symmetry reason and in fact is expected to be generated on 
the brane. Inclusion of this term often dramatically modifies the 
situation (e.g. can induce 4-dimensional gravity even if extra space is 
5-dimensional Minkowskian \cite{gia}). Thus, it would be interesting to explore
dynamical scenarios without fixing the shape of the brane.


\begin{thebibliography}{99}
\bibitem{example} The prime example here
being the duality between type IIB on
AdS$_5\times S^5$ and ${\cal N}=4$ supersymmetric $U(N)$
Yang-Mills in $d=4$ with
coupling $g_{YM}$ (the t'Hooft coupling is defined as
$\lambda=g_{YM}^2N$). See \cite{malda}.
\bibitem{o} Notice that the influence of the KK modes should be
important near the big-bang/big-crunch, but more or less negligible in the
middle of the evolution.
\bibitem{aclaracion}
Although $H^2$ is not a compact space, the low energy degrees of freedom 
cannot be excited on this manifold because they
have much larger wave-lengths when compared to the characteristic length of $H^2$.
\end{thebibliography}
\end{document}